\documentclass[conference]{IEEEtran}
\IEEEoverridecommandlockouts
\usepackage{cite}
\usepackage{amsmath,amssymb,amsfonts}
\usepackage{algorithmic}
\usepackage{graphicx}
\usepackage{textcomp}
\usepackage{xcolor}
\usepackage{algorithm}
\usepackage{algorithmic}
\usepackage{soul}
\usepackage{booktabs}
\def\BibTeX{{\rm B\kern-.05em{\sc i\kern-.025em b}\kern-.08em T\kern-.1667em\lower.7ex\hbox{E}\kern-.125emX}}
\begin{document}

\title{Exploring Conversational Agents as an Effective Tool for Measuring Cognitive Biases in Decision-Making\\
}

\author{\IEEEauthorblockN{Stephen Pilli}
\IEEEauthorblockA{\textit{UCD School of Computer Science} \\
\textit{University College Dublin}\\
Dublin, Ireland \\
stephen.pilli@ucdconnect.ie}
}
\maketitle

\begin{abstract}
Heuristics and cognitive biases are an integral part of human decision-making. 
Automatically detecting a particular cognitive bias could enable intelligent tools to provide better decision support. 
Detecting the presence of a cognitive bias requires a hand-crafted experiment and human interpretation.
Our research aims to explore conversational agents as an effective tool to measure various cognitive biases in different domains. 
Our proposed conversational agent incorporates a bias measurement mechanism that is informed by the existing experimental designs and various experimental tasks identified in the literature.
Our initial experiments to measure framing and loss-aversion biases indicate that conversational agents can be effectively used to measure the biases. 
\end{abstract}

\begin{IEEEkeywords}
cognitive biases, decision-making, conversational agents, bias measurement
\end{IEEEkeywords}

\section{Introduction}
Almost all animals (including humans) use heuristics in their daily decision-making processes, which have been shown to be surprisingly effective~\cite{gigerenzer_homo_2009}. 
Heuristics are characterised by the frugality of information and computation required to reach a satisfactory solution. 
While the near-universal prevalence of heuristics is testament to their evolutionary 'staying power', heuristics have also been implicated in various sub-optimal (or even wrong) decisions made by human beings~\cite{kahneman_model_2005}.
These have been categorised as cognitive biases, and more than two hundred have been catalogued, with different kinds of decisions being affected by different biases~\cite{carterBehavioralSupplyManagement2007}. 
Behaviour change techniques like nudges can leverage biases to tailor effective interventions~\cite{weinmann2016a}~\cite{mirsch2017a}.
Unfortunately, most bias measurement mechanisms require human interpretation of carefully designed experiments. 
The advent of Big Data and smartphones has added a new dimension to measuring bias. Not only are smartphones ubiquitous, but humans are now accustomed to using natural language as a medium to interact with them (e.g., Siri and Google Assistant). 
This means many contextual machine-human conversations could be used to measure bias. 
To the best of our knowledge, there is very little work on approaches to automate bias measurement in literature. We posit that conversational agents can be used to measure (and possibly ameliorate) bias among individuals computationally. Human-machine conversations can be engineered to provide information about the human interlocutor's thought process and clues about the possible biases they exhibit.

Although numerous cognitive biases have been identified through various studies, not all humans are susceptible to the same sub-set of biases. 
For example, some individuals may be affected by the framing effect, which refers to a cognitive bias in which individuals make choices influenced by how options are framed, whether they are presented with a positive or negative tone~\cite{recasens2013linguistic}.
Similarly, loss-aversion bias may affect others, which is the tendency to prefer avoiding losses over acquiring equivalent gains~\cite{ert2013descriptive}.
The approaches employed in the existing literature use carefully crafted experiments to measure these biases. 
Automated methods for measuring these biases are crucial in behaviour change techniques like nudging, yet this area remains unexplored. 

Our research revealed that existing experimental tasks for cognitive bias measurement often necessitate multiple interactions or strategic elicitation to comprehensively evaluate users' biases. However, we failed to identify a standard approach to measure cognitive bias using digital technologies. 

The decisions made by the individual on an experimental task could be influenced by factors other than cognitive bias. Therefore, one-shot (single measure) cognitive tasks are prone to classifying an unsystematic error with a systematic one. Repeated measures of experimental design can address this problem. We have identified that conversational agents can naturally facilitate repeated measures through multi-turn interactions. 

Moreover, our study highlighted that conversational agents have not been potentially used to measure cognitive biases. Conversational agents offer a richer experience by incorporating a conversational user interface (CUI) compared to the currently prevalent graphical user interfaces.

\section{Related Work}
The existing research on cognitive bias measurement is relatively limited. 
Within the digital technologies context, we have identified two significant areas where cognitive bias measurement has been studied. The first area pertains to decision-making, while the second is related to digital nudging. 
In decision-making, researchers have explored how cognitive biases manifest in digital environments and influence individuals' choices and judgements.  
We have identified various articles that conceptualised detecting cognitive biases in decision-making. For instance, MacShane et al. presented the OntoAgent model that supports the automatic detection of decision-making biases in clinical medicine setting~\cite{mcshane2013modeling}. Gulati et al. proposed a unifying, task-independent AI-based framework to automatically identify cognitive biases from observed human behaviour~\cite{gulati2022biased}. 

In a similar manner, various articles ~\cite{jung2018} ~\cite{stryja2019digital} ~\cite{rieger2020toward}~\cite{seeber2020machines} have proposed a conceptual approach to measure and mitigate biases in various tasks and aid decision-making. 
Echterhoff et al.~\cite{echterhoff_ai-moderated_2022} studied anchoring bias in decision-makers, like college admission officers, who are influenced by their recent decisions. They proposed an algorithm to detect existing anchored decisions, minimise sequential dependencies, and alleviate decision inaccuracies.
Reicherts et al. ~\cite{reicherts2022extending} developed a chatbot to help users make better investment decisions by triggering critical thinking  
by mitigating hindsight bias, disposition effect, and recency effect.

The second area is digital nudging, a concept that involves the usage of digital interfaces to guide user behaviour subtly. 
Determining heuristics and biases that might be at play in delivering a suitable intervention is crucial a stage in digital nudging
~\cite{weinmann2016digital}~\cite{schneider2018digital}~\cite{mirsch2017digital}~\cite{mills2022personalized}. 

Barev et al. ~\cite{barev2021dark} presented an experimental task to measure and record responses to the messages that use the framing effect. 
A user study was provided on how a privacy message that employs the framing effect in a digital work environment successfully lowers users' desire to expose personal information, changes their sense of threat, and affects negative emotions. 

Gena et al. ~\cite{gena2019personalization} 
compared three persuasive scenarios: Argumentum Ad Populum fallacy, Group-Ad Populum fallacy, and a neutral condition with no fallacy. 

To evaluate individuals' tendencies towards cognitive biases, the researchers Stryja et al. ~\cite{stryja2019digital} 
investigated loss aversion and the omission of action bias in investment decision-making. 

Momsen et al. ~\cite{momsen2014intention} presents findings from an original survey experiment investigating the impact of different interventions on individuals' choices between renewable and conventional energy contracts. 

Our review of the existing works identified notable gaps. First, a standard framework was not identified that can enable researchers to employ suitable types of cognitive tasks to incorporate into technologies that can be used to measure cognitive biases. Second, experimental tasks of existing cognitive bias experiments often require multiple interactions or strategic elicitation to assess users' biases. While CUIs can facilitate such interactions, the existing literature predominantly relies on graphical user interfaces as the primary medium for cognitive bias measurement.

\section{Research Questions}

\subsection{Cognitive Bias Measurement}

Each cognitive bias requires a unique experimental approach across different application domains, leading to various experimental tasks (decision tasks) incorporated into an experimental design. A single cognitive bias can be measured using various types of decision tasks. 
Therefore, it is essential to identify a prominent decision task among the empirical articles identified. These identified tasks can be used to abstract a representative.
Additionally, identifying commonalities among these representative decision task types will aid in establishing guiding principles in selecting appropriate decision tasks for their experiments. Our research question addresses these gaps to develop a more systematic approach.

\begin{itemize}
    \item \textbf{\textit{RQ1}}\label{label:rq1} How are cognitive biases measured in the decision-making and digital-nudging literature?
    \begin{itemize}
        \item \textit{RQ 1.1}\label{label:rq1.1} What are the existing representative decision tasks employed in cognitive bias measurement experiments?
        \item \textit{RQ 1.2}\label{label:rq1.2} What are the commonalities among the representative decision tasks?
    \end{itemize}
 \end{itemize}

\subsection{Conversational Agents to Measure Cognitive Biases}
A decision task (experimental task in the context of decision-making) involves presenting choices to the user, and their responses to these choices are recorded for subsequent analysis to measure biases. The analysis employs specific experimental design approaches, including repeated measures design or independent group design. We aim to explore the possibility of developing a conversational agent that can be used to measure cognitive biases. The proposed conversational agent uses standard task-oriented dialogue systems architecture ~\cite{ni2023recent}. We modify the  Natural Language Generation (NLG) unit, which essentially uses a traditional (template or handcrafted) ~\cite{santhanam2019survey} or contemporary approaches (LLMs)~\cite{zheng2021stylized}~\cite{wei2023leveraging} to generate decision tasks as utterance. The system can analyse the responses of the users for these decision tasks. The statistical analysis reveals the presence of cognitive bias that we intended to measure. We formulate the research questions $RQ2.1$ and $RQ2.2$ to determine if conversational agents can be used to measure cognitive biases in various domains. Additionally, using $RQ2.3$ to $RQ2.5$, we effectively plan to evaluate the proposed system on various dimensions.
 \begin{itemize}
 	\item \textbf{\textit{RQ2}}\label{label:rq2}  How can cognitive biases be measured using conversational agents in various domains?
     \begin{itemize}
     \item \textit{RQ 2.1}\label{label:rq2.1} Which cognitive biases are most prominently exhibited by users during interactions with conversational agents across multiple domains?

     \item \textit{RQ 2.2}\label{label:rq2.2} How do the conversational agents differ in their ability to measure cognitive biases across multiple domains?
    
    
     \item \textit{RQ 2.3}\label{label:rq2.3} How does the confidence in the measurement of cognitive bias differ between repeated measures and a single measure?
    
     \item \textit{RQ 2.4}\label{label:rq2.4} How can two different cognitive biases be effectively measured within the context of the same dialogue?
    
     \item \textit{RQ 2.5}\label{label:rq2.5} How does the significance of cognitive bias measurements change when the dialogue involves high levels of information overload?
        \end{itemize}
 \end{itemize}

\section{Research Methodology}
Our extensive literature review revealed data that can potentially be used to address our first research question. 
The data consists of articles that employed an experimental approach to measure cognitive biases. Each of these experiments discusses a task (cognitive task or decision task).
By thoroughly examining these articles, we identified decision tasks that are utilised in the primary studies. For each cognitive bias, we abstracted a representative task. Further, we identified the principal commonalities among these decision tasks. The classifications from this analysis offered us valuable insights with broad applicability, making them relevant for addressing our second research question.

To address the second research question \textbf{\textit{RQ2}}, 
experimental design methodology is employed. These research questions entail investigating the feasibility of measuring cognitive biases in various domains (e.g., travel, finance, and fitness) using conversational agents. First, we aim to develop a conversational agent that can be used to engage a user on a specific task like flight booking or fitness planning. The standard architecture informs the design of this chatbot of task-oriented dialogue systems. Our proposed chatbot employs one repeated measures design. The strategies we developed while addressing our first research question guided the decision task design for each cognitive bias. The existing datasets can inform the dialogue flow or the conversational task design. We plan to recruit participants from Amazon Mechanical Turk to evaluate our chatbot design. The chatbot dialogue with the participants is recorded. A two-factor repeated measures experimental design will be employed to validate the findings of our chatbot. Here, we recruit participants for our control group. By statistically comparing the experimental group with the control group, we establish evidence for our research questions.

\section{Preliminary Experiments}

\begin{figure*}[htpb]
    \centering
    \includegraphics[width=120mm,scale=1.0]{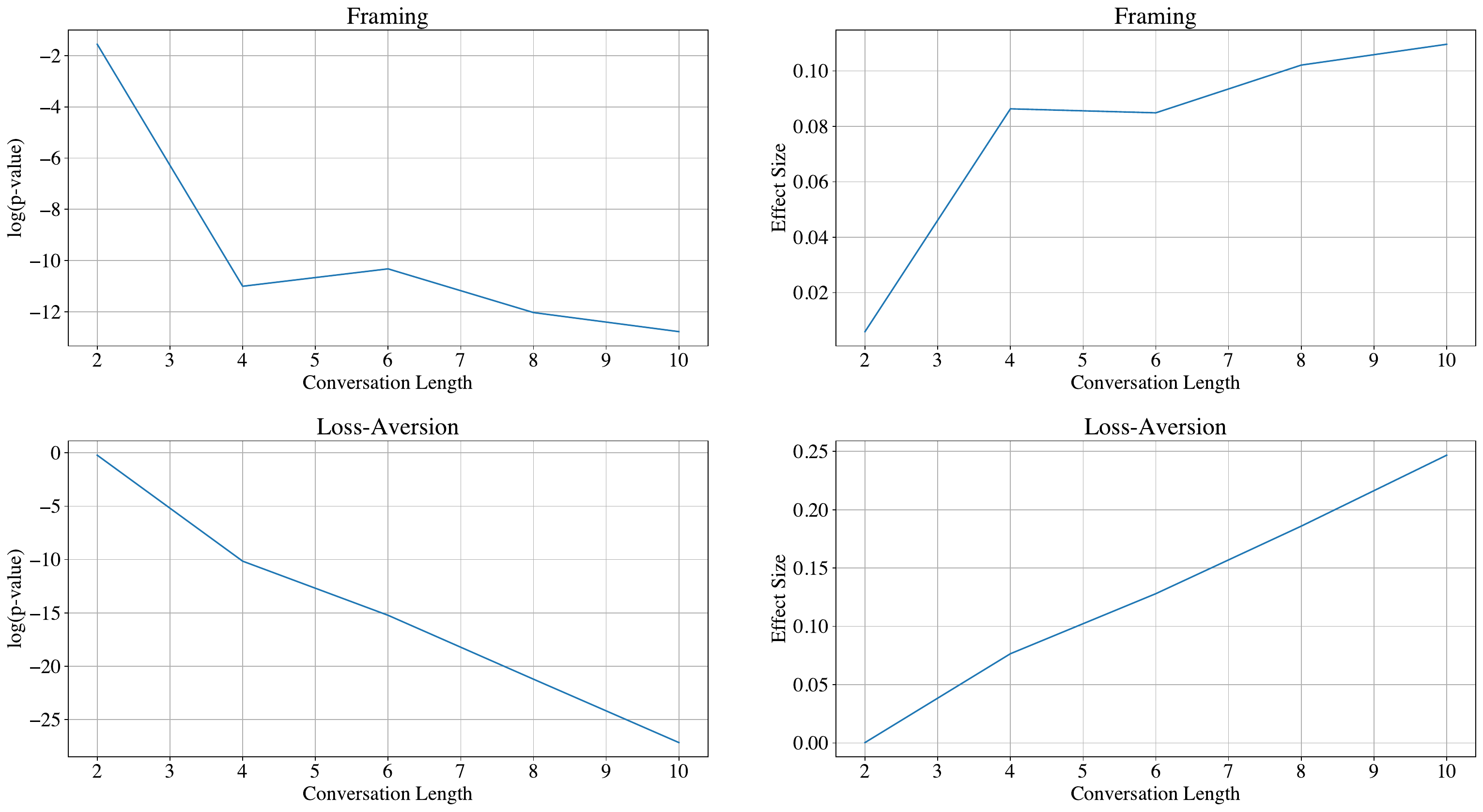}
    \caption{Conversation Length Vs \textit{p-value} and \textit{effect-size}.}
    \label{fig:convlen_stats}
\end{figure*}


Our preliminary experiment measures framing and loss-aversion biases in the travel domain. We formulated the following hypotheses to address our research questions: $RQ2.3$ - $RQ2.5$ specifically for an intended domain and two well-known cognitive biases.

\begin{enumerate}
    \item[\textbf{\textit{H1:}}]  \textit{A task-oriented chatbot can be used to measure framing and loss-aversion biases that involve decision-making through dialogue}.
    \item[\textbf{\textit{H2:}}] \textit{A task-oriented chatbot can be used to measure biases confidently within a few turns.}
\end{enumerate}

\subsection{Conversational Task - Trip Planning}
The chatbot is designed to engage users in a trip-planning conversation. 
The Frames dataset ~\cite{asri2017frames} consists of dialogues related to flight booking in the travel domain. These dialogues inform our task design in the travel domain. 
Our chatbot recommends flights to desired destinations, low-cost hotels to stay at, a local event to attend, or a restaurant to try. The chatbot pre-optimises the route for the flight costs. Therefore, the recommendations given by the chatbot are always cost-effective. Any recommendations given by the chatbot have two options. Therefore, a recommendation is a decision task. 

\begin{itemize}
    \item [\Large \textbf{(1)} \Large] Great! Let's start. Here are the first two cities. \textbf{Brussels} \textit{offers spectacular natural views}. However, the estimated carbon emissions for the trip to \textbf{Brussels} is higher than the trip to \textbf{Malaga}, \textit{a city with an area of 66 sq. km}. Where do you want to go?
\end{itemize}

In example $(1)$, considering carbon emissions between the two options, the flight to Malaga is the optimal choice while Brussels is sub-optimal. The chatbot creates a trip itinerary by repeatedly providing comparisons between two cities, hotels, and restaurants.
Various other factors can influence the decisions made by the user. Therefore, confounding variables like familiarity with the cities and entities have been controlled. 
The conversational task is \textit{ten} turns long.

\subsection{Experimental Tasks - Framing and Loss-aversion decision tasks}
A statement can use facts to convey a message. The same message can be conveyed more effectively using modifiers such as intensifying adjectives and phrases~\cite{recasens2013linguistic}.
The chatbot uses such modifiers while making recommendations to the user. In example (1), the chatbot uses an \textit{intensifying phrase} after Brussels and a \textit{fact phrase} after Malaga. 
A user that makes the sub-optimal choice (which results in higher estimated carbon emissions) over the optimal choice indicates susceptibility to framing bias. 
During the conversation, the chatbot makes five recommendations that can reveal framing bias. 
Nicolau et al. ~\cite{nicolau2012asymmetric} and Nguyen et al. ~\cite{nguyen2016linking} have shown loss-aversion during product choices and prices, including hotel booking in the tourism domain. Our chatbot is designed to utter these decision tasks as part of trip planning. 

\begin{itemize}
        \item[\large \textbf{(2)} \large]
    Great! Let's start. Here are the first two cities. Brussels is a city which has an area of 64 sqkm. However, the estimated carbon emissions for the trip to Brussels are higher than the trip to Malaga, a city which has an area of 66 sqkm. Where do you want to go? 
\end{itemize}

The chatbot design for the control group is similar to the experimental group in terms of functionality. However, the options presented to the control group participants in a decision task utterance (example $(1)$ for framing) will only have facts, and intensifying phrases will not be used.

\subsection{Experiments and Results}

We recruited participants from Amazon Mechanical Turk (AMT). To maintain the experiment's validity, it was ensured that the workers who participated in the control group  $(n=100)$ were different from those who participated in the experimental group $(n=100)$. The sample was diverse and comprised of various non-European individuals. Independent group experimental design was employed. 
The choices made by the participants are statistically analysed. A Mann-Whitney U Test determined statistically different between the two groups based on the participant's choices. 

\begin{table}[]
\caption{Results of the experiment on each bias}
\begin{center}
\begin{tabular}{llll}
                       & \textbf{Bias Found?} & \textbf{p-value} & \textbf{Effect Size} \\ \midrule
\textbf{Framing}       & \textit{yes}         & 0.001            & 0.109                \\ 
\textbf{Loss-Aversion} & \textit{yes}         & 0.001            & 0.205                \\ \bottomrule
\end{tabular}
\label{tbl:bias_pvalues_size}
\end{center}
\end{table}

Statistical analysis identified significant differences between the experimental and the control groups framing and loss-aversion biases with a small to medium effect size. Table~\ref{tbl:bias_pvalues_size} presents the effect size and the p-value. 
Our results confirm existing framing and loss aversion evidence in a one-shot setting. There has been criticism that unsystematic errors (random errors) are often attributed to systematic errors (biases)~\cite{gigerenzer2018bias}. This criticism often arises in the context of one-shot experiments.
The results in Figure ~\ref{fig:convlen_stats} provide evidence that the chatbot requires more than one turn to measure biases. The confidence (factor of effect size and p-value) in detecting biases increased with the increase in the length of the conversation. This evidence supports our hypothesis $H2$. 

\section{Conclusion and Future Work}
The results of our preliminary experiments provide valuable insights into the potential use of chatbots for measuring cognitive biases. By employing handcrafted decision-task utterances, we have laid the foundation for future research, which focuses on developing a model that can automatically generate decision-task utterances. Integrating such a model into an NLG module could streamline the entire measurement pipeline.

Our findings hold significant implications for the field of behavioural economics. In contrast to existing digital domain approaches that rely on one-shot (single measure) measurement of cognitive biases, our study demonstrates the advantages of repeated measures. Utilising more number of turns in the interaction process results in more robust and confident measurements of cognitive biases, addressing one of the significant challenges in traditional cognitive bias measurements~\cite{gigerenzer2018bias}. This approach offers a more concrete and reliable method for assessing these biases.

Our planned research aims to use conversational agents to strategically measure the top five prominent biases across three distinct domains. 
By doing so, we address the existing gap in the literature and establish conversational agents as an effective tool for measuring cognitive biases. The findings of our $RQ1$ will help researchers to make informed decisions on choosing task design for the bias intended to measure. 
The evidence from the planned experiment addressing $RQ2$ will act as a baseline for future researchers to develop fully automated conversational agents incorporating the advancements in NLU and NLG.

\section{Acknowledgements}
I would like to express my heartfelt gratitude to my supervisor, Dr. Vivek Nallur, and co-supervisor, Prof. Vincent Wade, for their invaluable guidance.
This work was conducted with the financial support of the Science Foundation Ireland Centre for Research Training in Digitally-Enhanced Reality (d-real) under Grant No. 18/CRT/6224. 
For the purpose of Open Access, the author has applied a CC BY public copyright licence to any Author Accepted Manuscript version arising from this submission.
The ethics of conducting research involving human participants is of utmost importance. This study was approved by the Institutional Review Board (IRB) and adheres to the ethical guidelines established by the declaration. Human Research Ethics Committee – Sciences (HREC-LS):  LS-LR-22-170-Pilli-Nallur.

\bibliographystyle{IEEEtran}
\bibliography{main.bib}
\end{document}